\documentstyle[prb,preprint,aps]{revtex}
\begin{document}
\draft
\title{Contact Injection into Polymer Light-Emitting Diodes}
\author{E. M. Conwell and M. W. Wu}
\address{Center for Photoinduced Charge Transfer, Chemistry Department,
University of Rochester, Rochester, New York 14627\\
and Xerox Corporation, Wilson Center for Technology, 114-22D, Webster,
New York 14580}
\maketitle

\begin{abstract}

The variation of current $I$ with voltage $V$ for poly(phenylene vinylene)
and other polymer light-emitting diodes has been attributed to
carriers tunneling into broad conduction and valence  bands.
In actuality the electrons and holes
tunnel into polaron levels and transport is by hopping among these levels.  
We show that for small injection the $I$-$V$ characteristic is determined
mainly by the image force, for large injection by space charge effects, 
but in both cases the strong variation of mobility with field due to disorder 
plays an important role.

\end{abstract}

\pacs{PACS: 73.40.Ns,72.20.Ht,71.20.Hk}

To improve the efficiency of polymer light-emitting diodes, LEDs, it is 
essential to understand and improve performance of the contacts.  Contact 
injection into LEDs made of MEH-PPV 
[poly(2-methoxy,5-(2$^\prime$-ethyl-hexoxy)-1,4-phenylene-vinylene]
has been attributed
to tunneling of the electrons and holes into a
broad conduction or valence band, 
respectively, in the polymer through interface barriers
arising from the band 
offset between the polymer and the metal electrodes.$^{1,2}$
Good agreement of
the $I$-$V$ characteristic with the field dependence given by Fowler-Nordheim
tunneling has been shown in references 1 and 2, although others did not 
find the Fowler-Nordheim field dependence for their diodes.

 The picture of carriers tunneling into a wide  band cannot  be 
correct for the conducting polymer samples now available.  First, 
because of the short conjugation lengths for typical polymers such 
as PPV, on average $\sim$ 6 or 7 monomers, continuum-like bands
become sets of discrete levels.  For the average conjugation length 
the level spacing is $\sim$ several $kT$ at  room temperature.$^{3,4}$
A more serious objection, however, is that injection of electrons or 
holes is into polaron levels,$^{5,6}$  which for electrons lie below
 the LUMO and for holes above the HOMO.  The separation of a 
polaron level from the LUMO or HOMO depends on the conjugation 
length of the segment.  For PPV calculations give the separation 
as 0.15 eV $^7$  or 0.2eV $^3$ for very long segments, increasing
to $\sim$ 0.7 eV for a 3 monomer long segment.  It is also a
consequence of the short conjugation lengths that theories 
based on the formation of a bipolaron lattice in the neighborhood 
of the contact cannot apply to currently available conducting 
polymer samples.\cite{saxx}

Based on the above considerations the scenario for conduction in an 
LED begins with a carrier from the metal tunneling into a polaron level 
close to  the contact.  The carrier  then diffuses and hops in the field 
from one conjugation length to another.  The variation 
in conjugation lengths, and the presence of defects, result in a 
spread in energy of the hopping sites (diagonal disorder).  An 
appropriate model for treating the transport in this system for 
low injection is the disorder model pioneered by B\"assler and 
associates.\cite{8} The distribution in energy of the polaron states
is taken as a Gaussian with variance $\sigma$.\cite{8}  From the
expected spread in conjugation lengths it is reasonable that
 $\sigma\sim 0.1$\ eV.\cite{con}

Critical for the behavior of a contact is the location in energy of the 
polaron states of the polymer relative to the Fermi  energy $E_F$ of the
metal.  For specificity we will discuss the case of electron injection 
into the polymer but the results apply to hole injection with the 
usual modifications. Internal  photoemission measurements, 
such as  those of Campbell {\em et al},$^5$
 yielding the energy required to inject an electron from a metal 
into a polymer, give the energy difference between $E_F$
and some average state in the polaron distribution.  We denote 
this energy  by $W$.  In what follows we calculate $I$-$V$
characteristics for a case of large $W$, which means small
injection, using the results of a Monte Carlo simulation based 
on the disorder model.\cite{9}  We then carry out a calculation
for $W\simeq 0$, which is the case for calcium contacts on
MEH-PPV, $^5$ using the classical approach of Rose\cite{10}
and Lampert.\cite{11}  In the latter case agreement is obtained,
for reasonable values of the parameters involved, with the $I$-$V$
characteristic of samples with only a  Ca contact injecting, 
providing we take into account the strong variation of mobility
 with field documented for PPV by Karg {\em et al}.\cite{12}

In the Monte Carlo simulation\cite{9} the energy $U$ of
a polaron site as a function of the perpendicular distance 
$x$ from the metal-polymer interface is written
\begin{equation}
U(x) = W - eEx - e^2/4\kappa x              
\end{equation}
Here $E$ is the electric field intensity and the last
term represents the 
image force, $\kappa$ being the dielectric constant.
In the presence of energetic 
disorder $U(x)$ gives the value of mean energy $\bar\varepsilon(x)$
of polaron sites 
at a distance $x$ from the interface.  Note that all energies are
measured 
relative to $E_F$.  With $W$ large, {\em e.g.} 0.6\ eV or greater,
injection is small and 
space charge may be neglected.  To calculate the incoming
 flux of carriers we 
have assumed that they tunnel into a polaron level 
with energy $\varepsilon$ at a distance $x$
from the interface at the rate $v_m(x) \exp(-\varepsilon/kT)$,
where $v_m(x)$ is a
distance-dependent prefactor.  With this assumption
the total number of carriers 
tunneling into polaron levels per second is
$v_m(x) \exp [(-\bar\varepsilon(x)+\sigma^2/2kT)/kT]$.
Thus the energy distribution 
of initially populated sites is displaced by $\sigma^2/kT$
from the available site 
distribution. By Monte Carlo simulation we followed the hopping
of the carriers through a sample of 12 layers, $331\times 331$
sites/layer, finally obtaining the
yield, {\em i.e.}, the fraction of injected carriers that escape
the return to the  electrode and reach the opposite boundary 
of the sample.\cite{9}  Even in the absence of disorder the image
force results in the great majority of carriers returning to the 
electrode at low fields.  An analytic solution for the ordered 
case based on Eq.\ (1) (neglecting space charge)\cite{13,14}
gives the yield as 0.3\% in a field
of $1.25\times 10^5$\ V/cm at 300\ K for
the parameters of Ref.\ \onlinecite{9}.   A field $\sim$ 10 times as
large is required for essentially
complete collection. The results of the analytic solution were in good
agreement with the Monte Carlo simulation for $\sigma=0$, indicating
that the simulation sample was thick enough.\cite{9}
In the disordered case the carriers have to 
overcome the random barriers due to disorder as well as the 
image force.  The result is that at fields of $\sim$ 10$^5$\ V/cm
mild disorder ($\sigma = 0.08$\ eV) results in a yield smaller
by a couple of orders of magnitude even though the injection
is larger.  In addition the yield increases more strongly with
 increasing field.\cite{9}  Current  {\em vs}. field may be obtained by
multiplying the yield  by the number of carriers entering the 
polymer per second. As shown in Fig.\ 1, for $W = 0.7$\ eV
and a sample length of 120\ nm, the effects of the image force 
and moderate disorder are a current increase by a factor of 10$^5$
as the applied voltage goes from 1 to 20\ volts. 

The photoemission data of Ref.\ 5  give the Fermi level for Ca
lying above the polaron level in MEH-PPV by 0.05\ eV.  This 
suggests that Ca provides an ohmic contact on MEH-PPV,
meaning that it could supply the maximum required current, i.e., 
space-charge limited current.  Calculations of current vs 
voltage for ohmic contacts were carried out in Refs.\ \onlinecite{10}
and \onlinecite{11}.  The calculations are based on Poisson's equation:
\begin{equation}
(\kappa/e)(dE/dx) = ( n  -\bar n ) + (n_t  -\bar n_t)\;,                
\end{equation}
where $n$ and $n_t$ are the densities of free and trapped
electrons, respectively, 
and $\bar n$ and $\bar n_t$   their respective average values
 for the sample in 
thermal and electrical equilibrium with the contact (no applied voltage). 
 The 
equation for current density $J$ is simplified by neglecting the
diffusion terms 
as was  done in Ref.\ \onlinecite{11}.  In this case we include the field
dependence of mobility 
$\mu$ found for hopping in many disordered systems,
including holes in PPV,\cite{12}
to give for the steady current
\begin{equation}
J = n e \mu_0 e^{\alpha \sqrt{E}} E
\end{equation}
where $\mu_0$  is the zero-field mobility and $\alpha$
was taken as a parameter in 
the calculations.  With the simplification of neglecting 
diffusion current the 
boundary condition at the cathode interface is $E =0$
at $x = 0$.  The
other electrode is taken as non-injecting.

Although there is an estimate of the trap density in MEH-PPV,  
specifically a few times 10$^{16}$/cm$^3$,\cite{15} we
lack information about their location. We therefore 
carried out numerical integration of Eqs. (2) and (3) 
for trap-free ($n_t =\bar n_t = 0$) and all-traps-filled
($n_t = N_t$, the total trap density ) cases.  As discussed
by Lampert, the trap-filled case does not give the correct 
current at low fields.  Instead there is a voltage threshold 
for current flow because before voltage is applied there
is already unneutralized charge in the traps which
prevents the injection of additional charge at the 
electrodes.  When $N_t\gg\bar n$, as is likely to be the
case here, the current rises very steeply with voltage 
beyond the  threshold.   Nevertheless the trap-filled solution 
should be good at high enough fields.

In Fig.\ 2 we compare our calculated results with the 
experimental data of Parker for a Ca contact on MEH-PPV 
in the electrons-only  case, i.e., with the work function 
of the other electrode too low to contribute significant 
hole current (at least below 18\ V)$^1$. It is seen that
good fits can be obtained for both the trap-free and
trap-filled calculations, although, as anticipated,
there is no fit at the lower fields in the latter case. 
For the trap-free case the parameters for the fit shown are $\alpha =
8\times 10^{-3}$cm$^{1/2}$ /V$^{1/2}$  and  $\mu_0 = 5
\times 10^{-11}$\ cm$^2$/Vs.  For  the trap-filled
cases both solid and dashed lines correspond 
to $\mu_0 = 5\times10^{-9}$\ cm$^2$/Vs.  For the solid line
 the other parameters are 
$\alpha = 4 \times10^{-3}$\ cm$^{1/2}$/V$^{1/2}$
and $N_t -\bar n_t -\bar n = 6\times 10^{16}$ /cm$^3$,
while for the dashed line $\alpha = 4.5\times10^{-3}$ cm$^{1/2}$/V$^{1/2}$
and $N_t  -\bar n_t -\bar n = 10^{17}$ /cm$^3$.   Smaller $\mu_0$
 is required for the fit to the trap free case because 
traps are not present  to keep the
current down at a given field. The parameter
 values are reasonable, particularly those for the 
solid line in the trap-filled case.  Extrapolation 
of the $\mu$ {\em vs} $E$ data for holes of
Ref.\ \onlinecite{12} to $E=0$ yields $\mu_0=5\times10^{-9}$\ cm$^2$/Vs.
Electron mobility is thought to be considerably 
lower than hole mobility in PPV, the difference most
likely being due to deep traps (perhaps carbonyls)
for the electrons.  It is not unreasonable that with Ca 
contacts the deep traps are filled and electron 
mobility becomes comparable to hole mobility.  
Karg {\em et al}. obtain $\alpha = 6\times10^{-3}$ cm$^{1/2}$/V$^{1/2}$,
close to the $\alpha$ values obtained here.  Finally, since $\bar n$ is
expected to be small, $N_t -\bar n_t$ is close to the trap
density estimated by Campbell {\em et al} from the
magnitude of the initial  increase in capacitance with 
forward bias.\cite{15}  From the good fit at high
voltages we conclude that the current there is
space-charge limited current.  It does not vary 
as V$^2$ because of
 the strong field dependence of $\mu$. We note that
the importance of space charge effects has also been 
stressed, albeit within the Fowler-Nordheim model, 
by Davids {\em et al}.\cite{16}

In summary, we have pointed out that Fowler-Nordheim
tunneling can not describe contact injection into
currently available polymer samples because the 
short conjugation lengths mean they cannot have
broad bands.  Injection and transport involve
only polaron levels.  These have a spread in energy
due to the conjugation length variations and other
defects.  This disorder, even though relatively mild,
can decrease injection by a couple of orders of magnitude
and makes mobility highly field dependent.  Ca
contacts to MEH-PPV should be ohmic because $E_F$ lies
above the polaron level.  $I$-$V$ characteristics resulting
from injection at Ca contacts are well fitted by theory 
for space-charge limited current with reasonable values 
for the mobility and its variation with field, and the trap density.

We are grateful to Dr.Yu. Gartstein for valuable
discussions and a critical reading of the manuscript.
  We acknowledge the support of the National 
Science Foundation under Science and Technology
 Center grant CHE912001.

\begin{figure}
\caption{Current density vs voltage from Monte Carlo
 simulation for a disordered sample ($\sigma = 0.08$\ eV)
 with one injecting contact with large $W$ (0.7 eV).  Sample length is
120\ nm.}
\end{figure}

\begin{figure}
\caption{The dots represent $I$-$V$ data from Ref.\ 1 for 120\ nm
 long MEH-PPV LEDs with one  Ca contact and the other 
contact Mg ($\bullet$) or Nd (o).  The lines represent
theoretical fits to the data for the trap-free and trap-filled 
cases.}
\end{figure}

\end{document}